\documentclass[11pt]{article}
\pdfoutput=1
\usepackage{latexsym}
\usepackage{amssymb}
\usepackage{amsmath}
\usepackage{color}
\usepackage{graphicx}
\usepackage{slashed}

\newcommand{\lsim}{\lesssim}
\newcommand{\gsim}{\gtrsim}
\newcommand{\tr}{{\rm Tr}}

\begin{document}
\pagestyle{empty}

\begin{flushright}
KEK-TH-1828
\end{flushright}

\vspace{3cm}

\begin{center}

{\bf\LARGE Topology in QCD and the axion abundance} 
\\

\vspace*{1.5cm}
{\large 
Ryuichiro Kitano and Norikazu Yamada
} \\
\vspace*{0.5cm}

{\it
KEK Theory Center, Tsukuba 305-0801, Japan\\
Department of Particle and Nuclear Physics\\
The Graduate University for Advanced Studies (Sokendai)\\
Tsukuba 305-0801, Japan\\
}

\end{center}

\vspace*{1.0cm}

\begin{abstract}
\baselineskip=18pt
\setcounter{page}{2}
\pagestyle{plain}
\baselineskip=16pt
\pagestyle{plain}
{
%
The temperature dependence of the topological susceptibility in QCD,
$\chi_t$, essentially determines the abundance of the QCD axion in the
Universe, and is commonly estimated, based on the instanton picture, to
be a certain negative power of temperature.  While lattice QCD should be
able to check this behavior in principle, the temperature range where
lattice QCD works is rather limited in practice, because the topological
charge is apt to freezes at high temperatures.
In this work, two exploratory studies are presented.  In the first part,
we try to specify the temperature range in the quenched approximation.
Since our purpose here is to estimate the range expected in unquenched
QCD through quenched simulations, hybrid Monte Carlo (HMC) algorithm is
employed instead of heatbath algorithm.
We obtain an indication that unquenched calculations of $\chi_t$
encounter the serious problem of autocorrelation already at $T \sim
2\,T_c$ or even below with the plain HMC.
In the second part, we revisit the axion abundance.
The absolute value and the temperature dependence of $\chi_t$ in real
QCD can be significantly different from that in the quenched
approximation, and is not well established above the critical
temperature.  Motivated by this fact and precedent arguments which
disagree with the conventional instanton picture, we estimate the axion
abundance in an extreme case where $\chi_t$ decreases much faster than
the conventional power-like behavior.  We find a significant enhancement
of the axion abundance in such a case.  
}
\end{abstract} 

\newpage
\baselineskip=18pt
\setcounter{page}{2}
\pagestyle{plain}
\baselineskip=18pt
\pagestyle{plain}

\setcounter{footnote}{0}

\section{Introduction}

It is widely believed that the instanton calculus in
QCD~\cite{'tHooft:1976fv} makes sense at high temperatures. The
asymptotic freedom ensures the perturbative expansion sensible and, most
importantly, the infrared divergences from the large instanton
contributions are cut-off by the Debye length, providing finite results
after the integration over the instanton size~\cite{Pisarski:1980md}.

In the semi-classical instanton picture, physics becomes
$\theta$-parameter dependent by the instanton contributions to the path
integral~\cite{Callan:1977gz}. The instanton calculus indicates that
such a dependence is proportional to the product of the quark masses,
$m_q^{N_f}$ and $\Lambda_{\rm QCD}^b$, where $b$ is the beta-function
coefficient, $b = 11N_c/3-2N_f/3$.
For example, the topological susceptibility, $\chi_t = (\partial^2 /
\partial \theta^2) V_{\rm eff}(\theta)$, is proportional to $m_q^{N_f}
\Lambda_{\rm QCD}^b T^{4 - N_f - b}$ by the dimensional analysis.

In the QCD axion model to solve the strong CP problem, the $\theta$
angle is promoted to the axion field $a(x) / f_a$, where $f_a$ is the
axion decay constant~\cite{Peccei:1977hh, Peccei:1977ur,
Weinberg:1977ma, Wilczek:1982rv, Kim:1979if, Shifman:1979if,
Dine:1981rt, Zhitnitsky:1980tq}. The topological susceptibility is
directly related to the mass of the axion as $\chi_t = m_a^2
f_a^2$. Therefore, the temperature dependence of $\chi_t$ discussed
above represents that of the axion mass, which is important for the
calculation of the axion abundance in the Universe. In the misalignment
mechanism for the axion generation in the early Universe, the axion
number density is proportional to the axion mass at the temperature at
which the axion field starts coherent
oscillations~\cite{Preskill:1982cy, Abbott:1982af, Dine:1982ah}. The
instanton based estimation of the temperature dependence is commonly
used in the literature, and predicts that the axion can naturally be
dark matter of the Universe when $m_a \sim 10^{-5}$~{\rm eV}, whereas
the astrophysical bound on $m_a$ is $m_a \lesssim 10^{-2}$~{\rm
eV}.
(See, e.g.,~\cite{Engel:1990zd}.) The allowed region, $10^{-5}~{\rm eV}
\lesssim m_a \lesssim 10^{-2}~{\rm eV}$, is called the ``axion window.''

There have been arguments which imply the disappearance of the instanton
effects at high temperatures.
In Ref.~\cite{Cohen:1996ng}, it has been argued in two-flavor QCD that
the difference between the two-point functions of iso-singlet and
iso-vector scalar operators vanishes in the chiral limit as fast as
$O(m_q^2)$ when the temperature is higher than the critical temperature.
By the Ward-Takahashi identities, this immediately means that $\chi_t$
should be at most an $O(m_q^4)$ quantity, that is different from
$O(m_q^2)$ from the instanton calculus.
Moreover, in a recent paper~\cite{Aoki:2012yj}, it is claimed that
$\chi_t$ is $O(m_q^N)$ with an arbitrary $N$ and thus it is vanishing
even with finite quark masses.
Although these are somewhat surprising results, it is certainly possible
that the instanton picture fails to describe the full quantum mechanical
vacuum~\cite{Witten:1978bc}. If that is the case, the estimation of the
axion abundance is significantly affected.

The lattice simulations can in principle discriminate whether instantons
make sense or not. However, the current situation is not conclusive.
In Refs.~\cite{Buchoff:2013nra, Bhattacharya:2014ara}, the temperature
dependence of various susceptibilities are measured, which seem to be
consistent with the semi-classical instanton picture and but at the same
time not to exclude other possibilities.
Meanwhile,
the analysis in Ref.~\cite{Cossu:2013uua} suggests the effective
restoration of $U_A(1)$ symmetry right above the critical temperature,
indicating the disappearance of the instanton effects.
The presence or absence of the $U_A(1)$ symmetry at the critical
temperature $T_c$ can also be inferred through the nature of the chiral
phase transition of massless two flavor QCD as attempted in, e.g.,
Ref.~\cite{Ejiri:2012rr}
(For a related work analyzing the system with the renormalization group
flow, see Ref.~\cite{Sato:2014axa}).
See Ref.~\cite{Dine:2014dga} for a new proposal to extract the
instanton effects in lattice QCD, and Ref.~\cite{Hanada:2015gsa} for the
approach from holography.

One unavoidable problem on the lattice is that the range of the
temperature in which one can reliably study the temperature dependence
of $\chi_t$ is rather limited, because the net susceptibility
$\chi_t V$, where $V$ is the volume, rapidly decreases with temperature
and hence the topology tends to freeze at high temperature.
This problem is fatal especially in dynamical lattice QCD simulations
since one can not realize arbitrary large volumes while keeping both the
light quark masses and lattice artifacts reasonably small.
Then, it is natural to ask to what temperature $\chi_t$ can be
reliably calculated in dynamical QCD with a typical setup.
From this viewpoint, determining the precise value of quenched
$\chi_t(T)$ with large volumes and huge statistics would not provide
useful information.

For the reliable calculation of $\chi_t$, it is preferable to use
lattice chiral fermion for dynamical quarks.
But such dynamical simulations at high temperatures are too costly to
start with.
As the second best, we choose to perform quenched simulations with
hybrid Monte Carlo (HMC) algorithm and Iwasaki gauge action.
The reasons for these choices are as follows.
To answer the above question, the simulation environment needs to be as
close to that in dynamical one as possible.
Long auto-correlation of the topological charge is one of the serious
bottlenecks in dynamical simulations, and the heatbath algorithm, which
is available only in the quenched approximation, is faster but has
totally different property from HMC in this regard.
Iwasaki gauge action is known to result in relatively long
autocorrelation time for the topological
charge~\cite{DeGrand:2002vu,Antonio:2006px}.
Thus, we expect that more information about dynamical simulations
will be gained from quenched one by taking HMC and Iwasaki gauge action
rather than taking heatbath and the standard plaquette gauge action.

Another issue to be addressed is the definition of topological charge
$Q$.
There is a subtlety in measuring $Q$ if one adopts the field theoretical
(or bosonic) definition.
This method requires a suitable number of coolings of gauge
configurations.
If one applies it too much or too less, one would miss the right value
of $Q$.
Even if one ceases the cooling adequately, $Q$ thus obtained is not an
integer, and in the worst case it may be right in the middle of two
integers, which can cause misidentification of $Q$.
Misidentifications are potentially dangerous, especially at high
temperatures ({\it i.e.} when $\langle Q^2 \rangle=\chi_t(T)\,V$ is
tiny), because it can significantly affect $\chi_t(T)$ even if it occurs
only rarely.
Importantly, it is not possible to know the right value of $Q$ without
comparing that obtained with the fermionic definition based on
Atiyah-Singer index theorem.
From this viewpoint, the use of the fermionic definition has an
advantage, although it is much more demanding than the bosonic one.
Studying the autocorrelation of $Q$ usually requires high statistics.
Thus, this choice of the definition seriously limits the lattice size to
small.

In this paper, we first explore to which temperature quenched
simulations using HMC and Iwasaki gauge action are able to obtain
reliable results for $\chi_t$, where the topological charge $Q$ in
each configuration is determined by the number of zero modes of
overlap Dirac operator.
Because of this choice, the lattice volumes are limited to relatively
small.
Our quenched calculations show that
$\chi_t$ is undetermined above $T=2\ T_c$ for $N_t=4$ and $T=1.5\ T_c$
for $N_t=6$, where $N_t$ is the number of lattice sites in the time
direction.

This observation suggests two possibilities.
One is that the result with $N_t=6$ is right and $\chi_t$ suddenly
decreases at $T \lsim 1.5\,T_c$.
In the language of the semi-classical instanton picture, this behavior
can happen if there is a sharp short-distance cut-off in the instanton
size parameter $\rho$ so that small instantons do not contribute to the
path integral. (See~\cite{Schafer:1996wv, Chu:1994xz} for the instanton
based models and lattice results suggesting this picture.)
If $\chi_t$ exhibits such a sharp fall off, the estimation of the
abundance of the QCD axion in the Universe may be significantly
affected.
However, this possibility appears to be unlikely because our value of
$\chi_t$ with $N_t=4$ is reasonably consistent with the previous
calculations to $T\sim1.75\,T_c$.
Another possibility is that, in the standard HMC with a fixed acceptance
ratio, the suppression of $\chi_t$ due to the long autocorrelation,
which increases with a lattice volume, overcomes the enhancement by
enlarging the volume.
Through the analysis of the autocorrelation, we confirm the latter to be
the case.

Outcome of the first part is that with HMC the reliable calculation of
$\chi_t$ is difficult for $T \gsim 2\,T_c$ even in the quenched
approximation.
Since quenched simulations are always easier than dynamical one,
the above outcome or even worse should hold for dynamical QCD.

The second part of the paper goes as follows.
We first emphasize that even in the quenched theory it is hard to
justify instanton calculus from the first principles because instanton
description consists of the perturbative expansion in terms of the
strong coupling constant $\alpha_s(T)$ and its prediction is only
reliable above $T=$ a few GeV.
The same is true or the situation is even worse for dynamical QCD.
Furthermore, there exist convincing arguments that the presence of
light dynamical quarks could drastically change $\chi_t$ above $T_c$ as
stated above.
Therefore, not much has been known about the unquenched theory, and
there are many degrees of freedom in the choice of $T$ dependence of
$\chi_t$ above $T_c$.

We revisit the axion abundance with some extreme but yet allowed
temperature dependence of $\chi_t$.
We find that the axion abundance is significantly enhanced compared to
the conventional estimates with non power-like behavior of $\chi_t$ and,
for some cases, the axion window is closed depending on how quickly
$\chi_t$ disappears.

\section{Topological susceptibility on the lattice}

In the continuum, the topological susceptibility, $\chi_t$, is defined as
\begin{align}
 \chi_t& = 
\left(
1 \over 64 \pi^2
\right)^2
\int d^4 x \langle 
\epsilon^{\mu \nu \rho \sigma} F_{\mu \nu}^a
 F_{\rho \sigma}^a (x)
\epsilon^{\mu' \nu' \rho' \sigma'} F_{\mu' \nu'}^b
 F_{\rho' \sigma'}^b (0) \rangle.
\label{eq:chit-def}
\end{align}
This quantity is related to the axion mass, $m_a$, through
\begin{align}
 \chi_t & = m_a^2 f_a^2,
\label{eq:axionmass}
\end{align}
where $f_a$ is the axion decay constant.
Through the Atiyah-Singer index theorem, one can also find
\begin{align}
 \chi_t& = {
\langle Q^2 \rangle \over V},
\label{eq:chitdef}
\end{align}
where the topological charge, $Q = n_+ - n_-$, is the difference between
the numbers of the left and the right-handed zero modes in the
eigenvectors of the Dirac operator.

On the lattice, the definition of gauge field strength tensor,
$F_{\mu\nu}$, is not unique, and so is $\chi_t$ if one simply follows
Eq.~(\ref{eq:chit-def}).  With the definitions of $F_{\mu\nu}$ commonly
used in the literature, $Q$ measured by
\begin{align}
 Q & = \left(
1 \over 64 \pi^2
\right)
\int d^4 x \epsilon^{\mu \nu \rho \sigma} F^a_{\mu \nu} F^a_{\rho \sigma},
\end{align}
takes non-integer values in general.
Thus, a technique called ``cooling'' is usually applied to guess the right
integer for $Q$.

Similarly to the continuum, an alternative is to count the zero
eigenvalues of the lattice Dirac operator.
This requires lattice Dirac operators to satisfy the Ginsparg-Wilson
relation~\cite{Ginsparg:1981bj}.
The overlap Dirac operator, $D_{\rm ov}$, is known as such an
operator~\cite{Neuberger:1998wv}, with which one can express $Q$ as
\begin{align}
 Q& = \tr \Gamma_5,
\label{eq:qdef}
\end{align}
where 
\begin{align}
    \Gamma_5
& = \gamma_5 \left(
1 - {D_{\rm ov} \over 2 M_0}
\right)
  = {1 \over 2} 
    \left[{\gamma_5} - {\rm sgn} (H_W(-M_0))
    \right].
\end{align}
Here, $H_W = \gamma_5 D_W(-M_0)$, and $D_W(-M_0)$ is the Wilson Dirac
operator with a negative mass parameter, $-M_0$.
This definition provides an integer value of
$Q$ in each configuration. The trace of the spinor indices of $\Gamma_5$,
${\rm tr} \Gamma_5$, can provide a definition of the local value of the
topological density, $(1/64 \pi^2) \epsilon^{\mu \nu \rho \sigma} F_{\mu
\nu}^a F_{\rho \sigma}^a$.

As stated in the introduction, the bosonic definition has the chance to
misidentify $Q$.
While, below and around the (pseudo-)critical temperature, previous
works have revealed that the bosonic and fermionic definitions of $Q$
give consistent results for $\chi_t$, it is not well established above
$1.5\, T_c$ yet.
Especially, we have to keep in mind that a tiny amount of
misidentification of $Q$ may bring significant effects to the resulting
$\chi_t$ at high temperatures as follows.

Whether the instanton picture are correct, it is certainly true that
$\chi_t$ decreases with the temperature and $\chi_t V$ as well.
Since $\chi_t V$ represents a net width of the fluctuation of $Q$, the
fraction of configurations with non-zero $Q$ becomes much smaller than
that with the vanishing $Q$ when $\chi_t V\ll 1$.
Suppose that $N_{\rm conf}$ configurations are generated on the lattice
volume of $N_V=N_s^3\times N_t$ and that $q$ of them belong to either
$Q=+1$ or $Q=-1$ while the others to $Q=0$.
This yields
  $a^4 \chi_t$
= $\langle Q^2 \rangle/ N_V$
= $(q + \delta_{\rm mis})/(N_{\rm conf}\ N_V)$, 
where $\delta_{\rm mis}/N_{\rm conf}$ represents the rate of
misidentification.
Then, it should be noted that, even if $\delta_{\rm mis}/N_{\rm conf}$ is tiny, 
$a^4 \chi_t$ may significantly deviate from a real value because the
fraction of nonzero $Q$ configurations, $q/N_{\rm conf}$, is also tiny
at high temperatures.
To avoid the misidentification, we adopt the fermionic definition in
Eq.~\eqref{eq:qdef} for the evaluation of $\chi_t$.

\section{$\chi_t(T)$ in the quenched approximation with HMC}
\subsection{lattice parameters}

To calculate the topological susceptibility at finite
temperatures, we perform lattice simulations of $SU(3)$ pure
Yang-Mills theory around and above $T_c$.
We employ the Hybrid Monte Carlo (HMC) algorithm to study the
autocorrelation of topological charge in HMC for the reason described in
introduction.
To avoid the {\em ordinary} finite size effects, the aspect ratio is set
to three or four, which is usually considered to be safe.
In addition, we have to recall that the size of fluctuation of $Q$ is
directly affected by a volume through $\langle Q^2 \rangle=\chi_t\,V$,
which is a finite size effect specific to and important in the
calculation of $\chi_t$.
The calculation is carried out on three lattice volumes using the
Iwasaki gauge action, $16^3\times 4$, $18^3\times 6$ and $24^3\times 6$.
The topological charge $Q$ is calculated at every ten trajectories.
The statistical errors are estimated by the standard jack-knife
method with the bin size of 500 trajectories.

The correspondence between $\beta$ and $T/T_c$ for the Iwasaki gauge
action is read from Ref.~\cite{Okamoto:1999hi}.
We accumulated 2000 to 40000 trajectories, depending on the simulation
parameters, which are summarized in Tab.~\ref{tab:simpara}.
\begin{table}[bt]
 \centering
 \begin{tabular}{c|ccrccl}
  $V/a^4$ & $\beta$ & $T/T_c$ & \# of traj. &
  $\delta\tau$ & acc. & \ \ $\chi_t/T_c^4$\\
  \hline
  $16^3 \times 4$ & 2.288 & 1.00 &  5000 & 1/55 & 0.79 & 0.126(12)\\
                  & 2.450 & 1.34 & 15000 & 1/60 & 0.78 & 0.176(17)$\times 10^{-1}$\\
                  & 2.522 & 1.50 & 19970 & 1/65 & 0.78 & 0.66(9)  $\times 10^{-2}$\\
                  & 2.623 & 1.75 & 39860 & 1/62 & 0.77 & 0.43(11) $\times 10^{-2}$\\
                  & 2.716 & 2.00 & 33660 & 1/67 & 0.79 & 0.90(41) $\times 10^{-3}$\\
                  & 2.802 & 2.25 & 43250 & 1/65 & 0.79 & [0.9(9)  $\times 10^{-4}$]\\
  \hline
  $18^3 \times 6$ & 2.445 & 0.89 &  4500 & 1/80 & 0.82 & 0.148(18)\\
                  & 2.515 & 1.00 & 19500 & 1/80 & 0.82 & 0.143(14)\\
                  & 2.710 & 1.34 & 17560 & 1/80 & 0.80 & [0.57(21)$\times 10^{-2}$]\\
                  & 2.794 & 1.50 & 15710 & 1/80 & 0.79 & \\
                  & 3.018 & 2.00 & 17050 & 1/80 & 0.78 & \\
  \hline
  $24^3 \times 6$ & 2.445 & 0.89 &  4600 & 1/77 & 0.73 & 0.111(15)\\
                  & 2.794 & 1.50 &  2140 & 1/100& 0.80 & [0.28(19) $\times 10^{-2}$]\\
                  & 3.018 & 2.00 &  4030 & 1/100& 0.78 &  \\
 \end{tabular}
 \caption{Simulation parameters. The trajectory length is always set to
 one. acc. denotes the acceptance ratio. Topological charge $Q$ is
 calculated every ten trajectories.
 The statistical errors are estimated by the jackknife method with the
 bin size of 500 trajectories.
 The blank indicates that $\chi_t$ can not be calculated due to too rare
 $Q$ changes.}
 \label{tab:simpara}
\end{table}

Lattice volumes chosen in the present work are somewhat smaller than
other pure YM calculations.
The reason is as follows.
In order to reduce the ambiguity associated with the definition of $Q$
as much as possible, we decided to use the fermionic definition for $Q$
to calculate $\chi_t$.
The purpose of this work is to find the highest temperature to which the
reliable extraction of $\chi_t$ is feasible in HMC, or in other words,
at which temperature the autocorrelation time becomes unreasonably
long.
Such a study usually requires high statistics.
Furthermore the fermionic definition is much more expensive than the
bosonic one.
Thus this choice of definition constrains lattice volumes to be small.

Here let us mention the difference of this work from
Ref.~\cite{Gattringer:2002mr}, which studies the $T$ dependence of
$\chi_t$ around the critical temperature, using the fermionic definition
for $Q$.  Two major differences are in the range of the temperature
explored and rigorous application of the overlap Dirac operator.  The
former is originating from the different motivation.  Let us explain
more about the latter.  In Ref.~\cite{Gattringer:2002mr}, the
identification of $Q$ is made using an approximate solution of the
Ginsparg-Wilson equation, while we stick to the exact one.  They differ
in important way.  To check the approximated solutions, in
Ref.~\cite{Gattringer:2002mr} the chirality of the zeromodes
($\psi_i^{\dag} \gamma_5 \psi_i$) is examined, and found that its
absolute value ranges from 0.4 to 0.9.  With our exact method, it only
takes +1 or -1 within a rounding error.  Therefore, in
Ref.~\cite{Gattringer:2002mr}, the possibility of the misidentification
cannot be excluded thoroughly.  Indeed, they observe that 2 to 9 \% of
configurations are misidentified by comparing their approximated
solutions and the exact one below the critical temperature.
Importantly, it is not clear how well the method based on the
approximation works at higher temperatures.  In our calculation, the
size of zeromodes and near-zeromodes of Hermitian overlap operator
$H_{\rm ov}$ differ by a factor of $O(10^{6})$, and we checked for all
zeromodes whether the complex conjugate pair is absent.  Thus, no
ambiguity exists.  While 2 to 9 \% of misidentifications will not
significantly affect $\chi_t$ at low temperatures, it does at high
temperatures as $\chi_t$ itself is very small there.  Note that both of
the two differences are essential in the study of axion.

\subsection{autocorrelation of $Q$}

\begin{figure}[bt]
\begin{center}
 \begin{tabular}{c}
  \includegraphics[width=0.9 \textwidth,bb=0 0 710 560,clip=true]
  {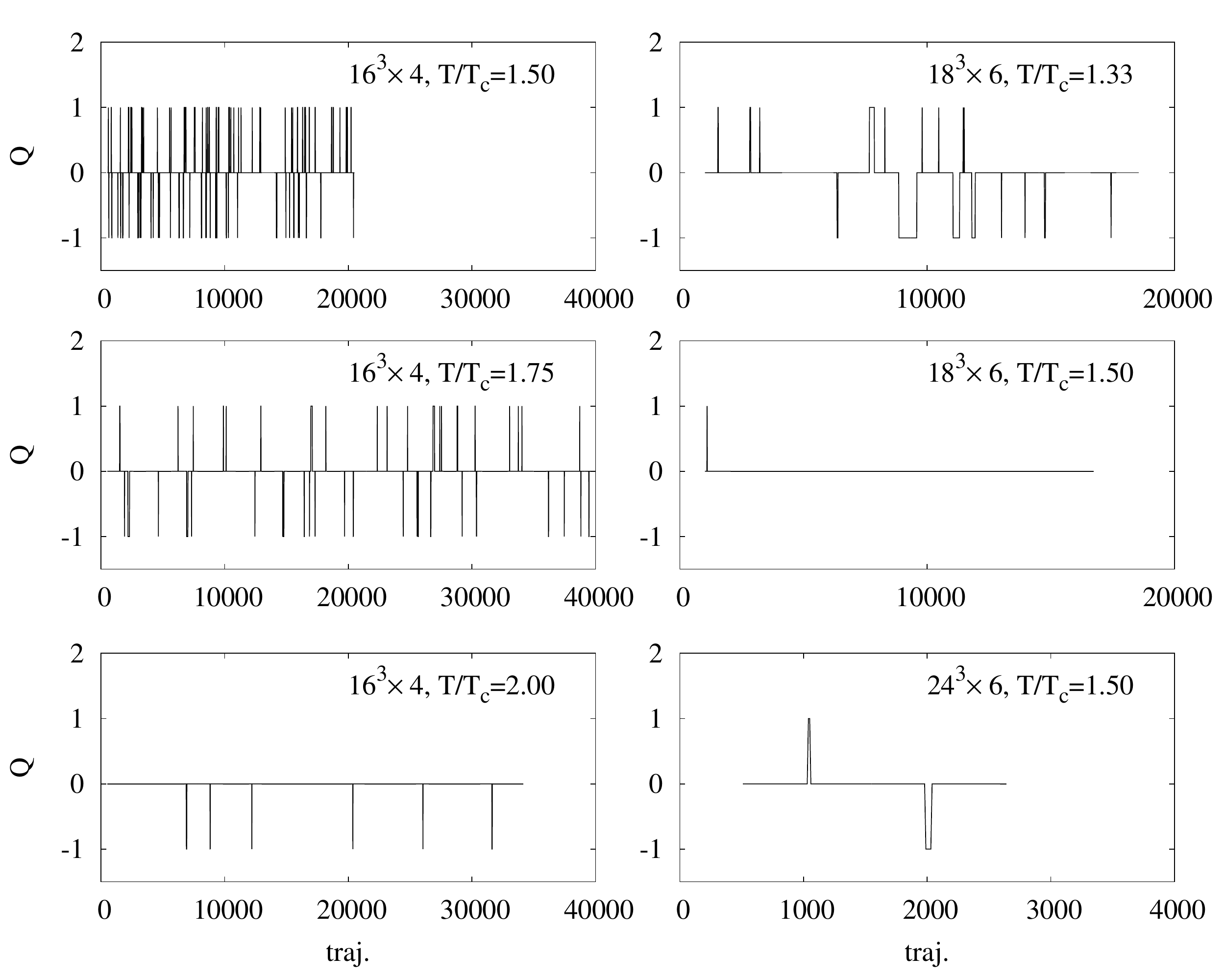}
 \end{tabular}
\end{center}
 \caption{History of topological charge $Q$ above $T_c$.
 }
\label{fig:history-all}
\end{figure}

Figure~\ref{fig:history-all} shows the history of $Q$ at several
simulations.
It is seen that, on our smallest lattices, $Q$ changes reasonably
frequently below $T\le 1.75\,T_c$.
But at $T=2\,T_c$, we only observe $Q=-1$ for nonzero $Q$.
Further increasing $T$ to $T=2.25\ T_c$, non-zero $Q$ was only observed
in one configuration in spite of our largest statistics.
Thus, $\chi_t$ is less reliable at $T \gsim 2\,T_c$.
Another remark, which is not plotted, is that while $|Q|$ fluctuates
over the range of $-8$ to $+8$ and $-2$ to $+2$ at $T=T_c$ and
$1.34\,T_c$, respectively, $Q$ only takes a value of $-1$, 0 or $+1$ at
$T=1.5\,T_c$ and $1.75\,T_c$.
While this is expected as $\chi_t$ decreases, the central values and the
statistical errors for those temperatures needs to be checked by
comparing other works.
(The comparison is made later in Fig.~\ref{fig:chiT}.)

It is naively expected that $Q$ fluctuates more frequently at larger
lattices.
However, the plots in Fig.~\ref{fig:history-all} shows the opposite
tendency, which indicates that the suppression of $\chi_t$ due to the
long autocorrelation in HMC with a fixed acceptance ratio overcomes the
enhancement by enlarging the volume for our choice of lattice setup.

\begin{figure}[htb]
\begin{center}
 \begin{tabular}{cc}
  \includegraphics[width=0.5 \textwidth,bb=0 0 360 240,clip=true]
  {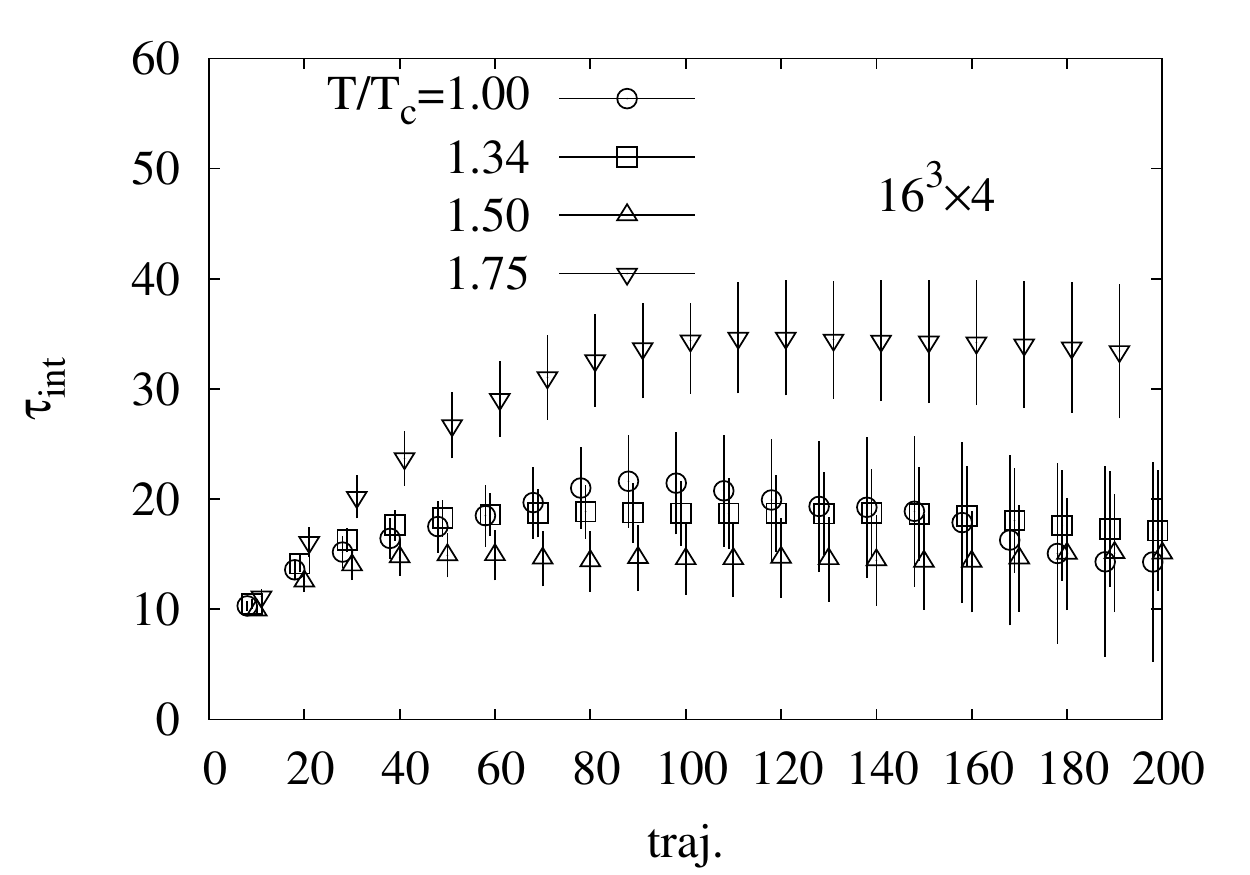}& \hspace{-4ex}
  \includegraphics[width=0.5 \textwidth,bb=0 0 360 240,clip=true]
  {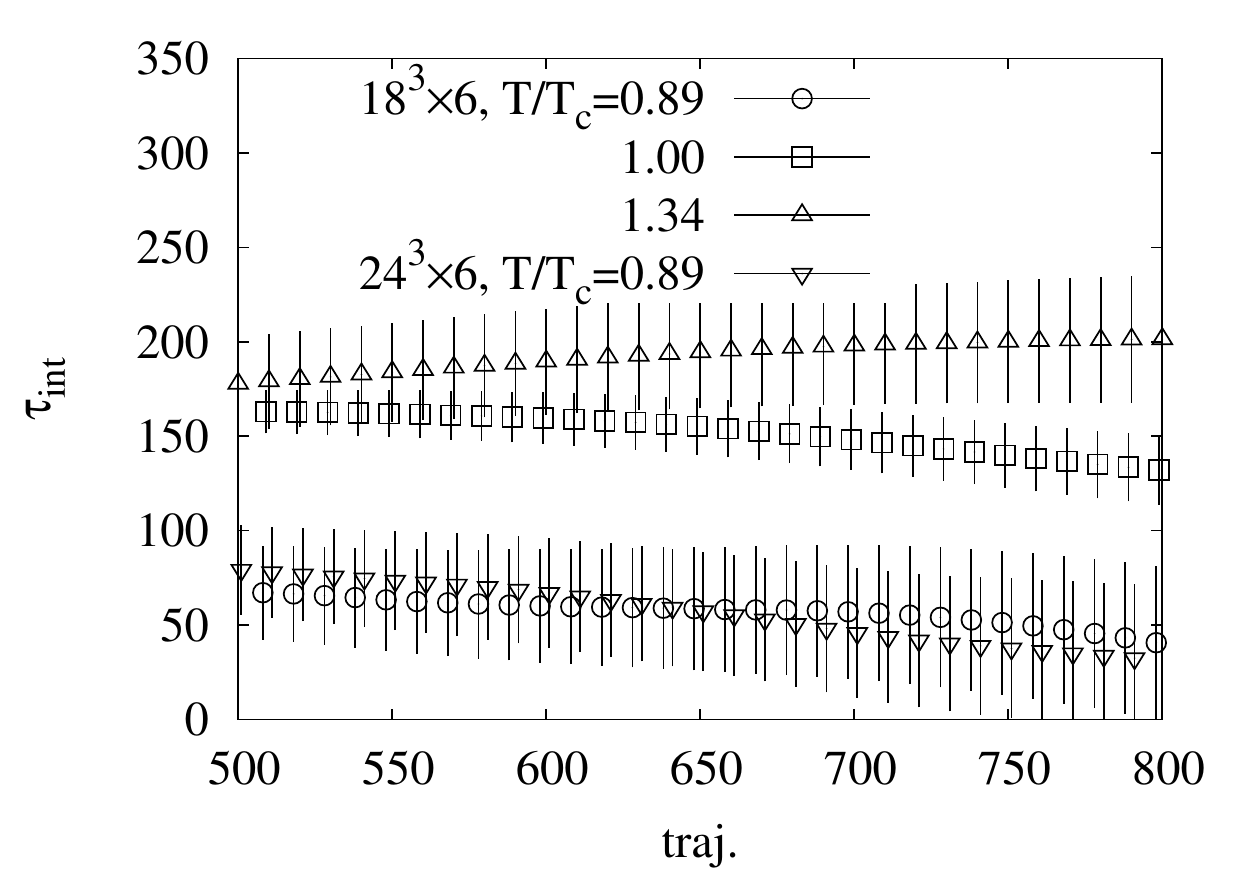}\\
 \end{tabular}
\end{center}
 \caption{Integrated autocorrelation time for $Q$ on $16^3\times 4$
 (left) and on the other larger lattices (right).
 }
\label{fig:autocorr-16x4}
\end{figure}
To quantify the autocorrelation, we estimate the integrated
autocorrelation time, $\tau_{\rm int}$.
Figure~\ref{fig:autocorr-16x4} shows that $\tau_{\rm int}$ takes 10 to 20
trajectories at $T\le 1.5\,T_c$ on $16^3\times 4$ and increases to
about 40 trajectories at $T\le 1.75\,T_c$.
Note that we could not obtain $\tau_{\rm int}$ at $T\ge 2.0\,T_c$.
We observe the qualitatively similar behavior for $18^3\times 6$
lattices except for the $T=T_c$ data, which shows $\tau_{\rm int}$
larger than other results obtained at $T\sim T_c$.

\subsection{$T$ dependence of $\chi_t$}

\begin{figure}[hbt]
\begin{center}
 \begin{tabular}{cc}
  \includegraphics[width=0.5 \textwidth ]
{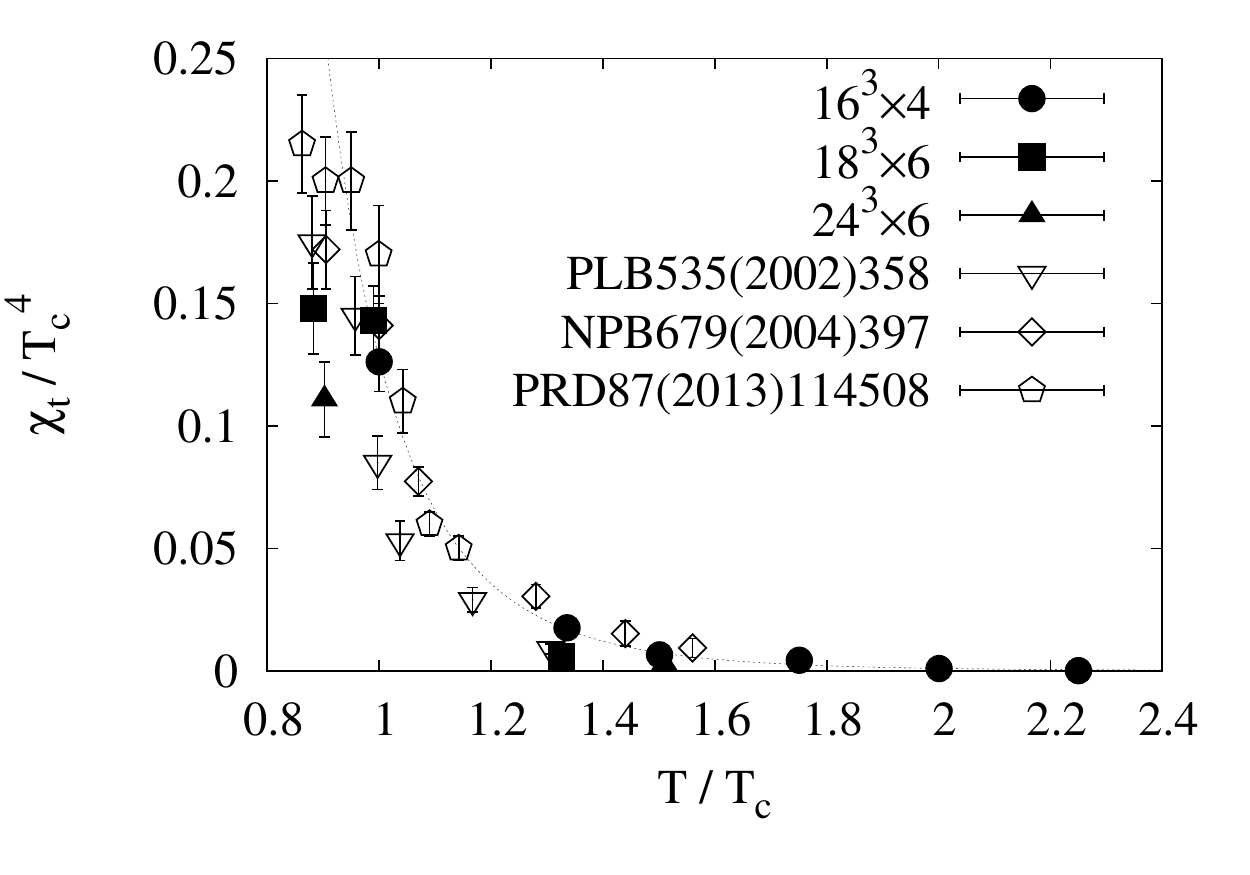} &
  \includegraphics[width=0.5 \textwidth ]
{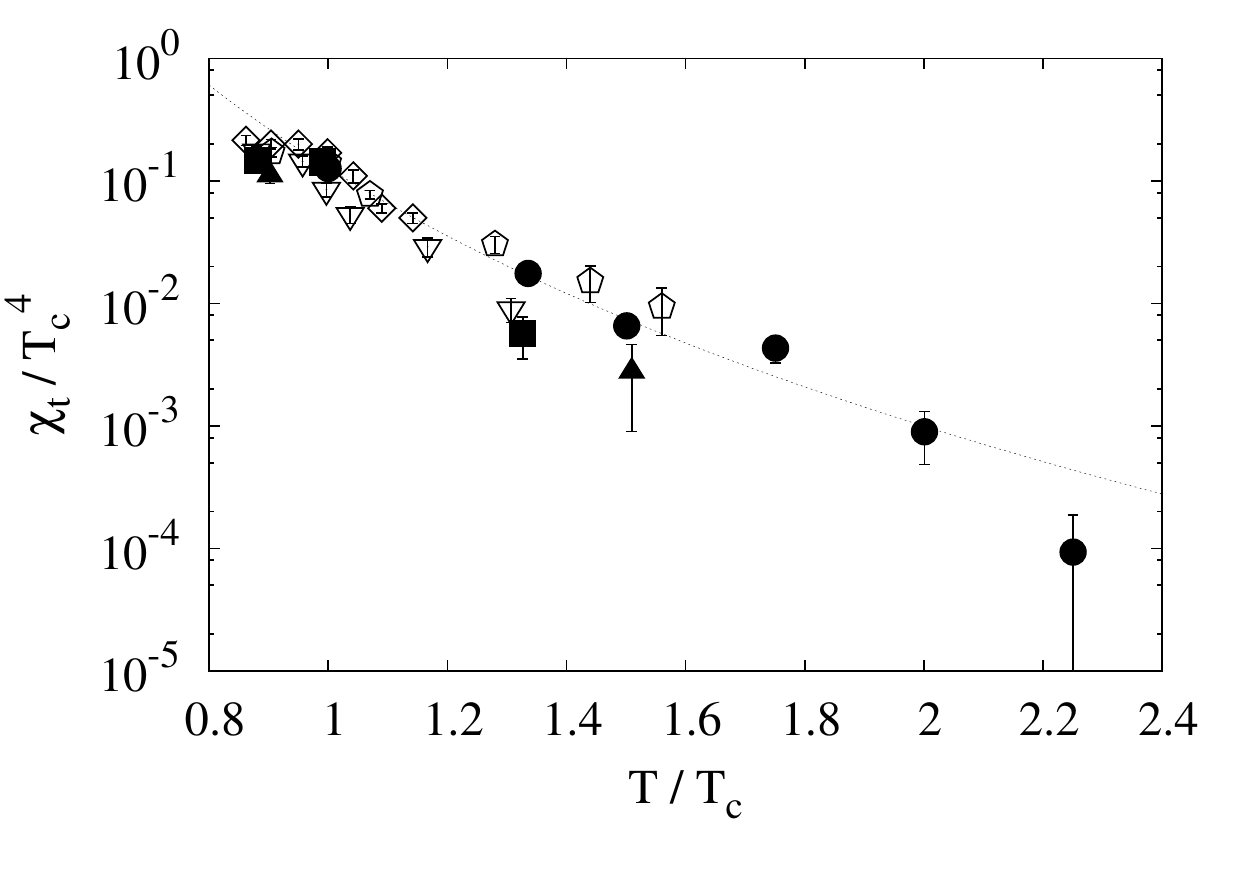} \\
 \end{tabular}
\end{center}
\caption{The temperature dependence of the topological susceptibility,
 $\chi_t$ in unit of $T_c^4$.
 The results of $16^3\times 4$ (filled circles), $18^3 \times 6$ (filled
 squares) and  $24^3\times 6$ (filled triangles) lattices with the
 standard HMC are shown.
 The dotted curve is the fit of five data points of $16^3\times 4$
 lattices obtained in the range $1\le T/T_c \le 2$.
 }
\label{fig:chiT}
\end{figure}
The figure~\ref{fig:chiT} shows the $T$ dependence of $\chi_t$.
It is seen that, restricting the data points to the one in which $Q$
shows a recognizable fluctuation, the results obtained from such
fluctuations are consistent with the existing results, for example,
given in Ref.~\cite{Gattringer:2002mr,Alles:1996nm,Bornyakov:2013iva}.
It is important to note that we have only observed $|Q|\le 1$ at
$T=1.50\,T_c$ and $1.75\,T_c$, but the resulting $\chi_t$s nevertheless
reasonably agree with the previous results.
It is also found that $\chi_t$ with different volumes shows consistency
within two standard deviations as long as the data showing a
recognizable fluctuation of $Q$ are concerned.

To see the consistency of our results with the instanton calculus, we
fit the five results in the range of $1\le T/T_c \le 2$ to the
form of $\propto 1/T^7$, which is suggested in the leading order
instanton calculus.
The fit quality is reasonable ($\chi^2/$d.o.f.= 0.97).

From these observations, we realized that with HMC it is difficult to
obtain the reliable $\chi_t$ above $T=2\ T_c$ or even $1.5\ T_c$
depending on the lattice volume, which immediately indicates the
difficulty in dynamical simulations.
Even if much faster computers were used, this upper bound will not
change significantly.
Thus, to estimate the $T$ dependence of $\chi_t$ at $O(10\,T_c)$, we
have to make a long extrapolation using those obtained in such a rather
limited range of $T$.
To push the limit upward as high as possible, it is crucial to explore
the HMC parameters or improve the algorithm.

Inspired by Refs.~\cite{Fukaya:2006vs,McGlynn:2013ava}, we tried, as an
attempt, to enhance the number of configurations with nonzero $Q$ by
inserting 
\begin{align}
 X = \det \left(
{H_W^2 + \mu^2 \over H_W^2 + \epsilon^2}
\right)^{N_\phi},
\end{align}
to the path integral, where $N_\phi$ is a positive integer.
Then, $\chi_t$ is calculated through
\begin{align}
 \chi_t& = {\langle (Q^2 / V) X X^{-1} \rangle 
\over
\langle X X^{-1} \rangle}
= 
{\langle (Q^2 / V) X^{-1} \rangle_X
\over
\langle X^{-1} \rangle_X},
\end{align}
where $\langle \cdots \rangle_X$ denotes the average over the
configurations generated with the extra reweighting factor $X$. 
For $\mu > \epsilon$, the insertion of $X$ enhances the eigenvalue
density in the small eigenvalue region, whereas eigenmodes with
eigenvalues $\lambda \gg \mu$ are left untouched.
Since, when the topology changes, the smallest eigenvalue of $H_W$ passes
through zero, the above factor is expected to increase such
opportunities.
However, after we performed some trial calculations, we realized that
this method does not always work and the fine tuning of $\mu$,
$\epsilon$ and $N_\phi$ are required.
Further investigations to improve the situation is in progress.

\section{Effects of dynamical quarks}

Let us discuss what would happen when we include the dynamical
quarks.
The naive guess would be that $\chi_t$ in the Yang-Milles theory is
multiplied by a factor of $m_q^{N_f} / \Lambda_{\rm QCD}^{N_f}$ since
$\chi_t$ should vanish when one of the quark masses goes to zero.

There can be more drastic possibilities.
If we accept the claims of the axial $U(1)$ restoration in two-flavor
QCD~\cite{Cohen:1996ng,Aoki:2012yj}, the $O(m_q^2)$ contributions to
$\chi_t$ is forbidden in two-flavor QCD.
Therefore, the possibility of just multiplying by
$m_q^{N_f} / \Lambda_{\rm QCD}^{N_f}$ is not consistent.
The results of Ref.~\cite{Aoki:2012yj} even forbid contributions with
any power of $m_q$ for a small $m_q$.
An extreme possibility one can consider is
\begin{align}
\chi_t (T) \sim \left \{ 
\begin{array}{ll}
 m_q \Lambda_{\rm QCD}^3, & T < T_c, \\
 m_q^{2} \Lambda_{\rm QCD}^{2} e^{-2 c(m_q) T^2 / T_c^2}, & T > T_c, \\
\end{array}
\right.
\label{eq:naivemodel2}
\end{align}
with $c(m_q) \to \infty$ as $m_q \to 0$, so that $\chi_t$ cannot be
expanded around $m_q = 0$.
Note that the results of Ref.~\cite{Aoki:2012yj} is contained as a
special case of eq.~(\ref{eq:naivemodel2}).
Since no unquenched result of $\chi_t$ is available at high
temperatures, we take $c(m_q)$ as a free parameter in the following
discussion.

\section{Axion abundance}

We discuss the impact on the axion abundance in the Universe for the
case where the dilute instanton gas approximation fails at high
temperatures.
One of the source of the axion energy density today is the coherent oscillations of
the axion field started in the early Universe. The equation of motion
for the axion field, $a$, is given by
\begin{align}
 {\ddot a} + 3 H \dot a = - m_a^2 (T) a,
\label{eq:axionEOM}
\end{align}
where $H$ is the Hubble parameter and $m_a(T)$ is the axion mass at a
temperature $T$. The axion mass is related to $\chi_t$ as in
Eq.~\eqref{eq:axionmass}.
This equation of motion leads to an oscillating solution.
Although the axion mass is temperature dependent, and
thus time dependent, the axion number density divided by $T^3$,
$n_a/T^3$, stays constant during the oscillation and its value is given
by
\begin{align}
 {n_a \over T^3}& \simeq {m_a(T_*) f_a^2 \theta^2 \over T_*^3},
\label{eq:axionND}
\end{align}
where $T_*$ is the temperature at which the axion field starts to
oscillate, and $\theta$ is the initial amplitude of $a/f_a$. In the
conventional estimates, one uses $m_a (T) \propto T^{-4}$ derived by the
instanton gas approximation. The temperature $T_*$ is obtained by
equating $m_a(T)$ and the Hubble parameter, $H(T) \sim T^2/M_{\rm Pl}$,
with $M_{\rm Pl}$ being the Planck scale, such as
\begin{align}
 T_*& \simeq 1~{\rm GeV} \cdot \left(
m_a \over 10^{-5}~{\rm eV}
\right)^{1/6}.
\label{eq:tempS}
\end{align}
By using this value and setting $m_a(T_*) \simeq 3 H(T_*)$, the energy
fraction today is finally given by
\begin{align}
 \Omega_a \simeq 0.2 
 \cdot \theta^2 \cdot \left(
m_a \over 10^{-5}~{\rm eV}
\right)^{-7/6},
\label{eq:conv}
\end{align}
for $\theta \ll 1$, and it is about a factor of two larger for $\theta
\sim O(1)$~\cite{Turner:1985si, Lyth:1991ub}.  According to this
formula, the axion can naturally be the dark matter of the Universe for
$m_a \sim 10^{-5}$~eV which corresponds to $f_a = 6 \times
10^{11}$~GeV. (See~Ref.~\cite{Wantz:2009it} for a recent detailed
calculation.)

If the axion mass $m_a(T)$ decreases much faster as discussed in the
previous section, the axion starts to oscillate near the critical
temperature $T_c \sim 150$~MeV independent of the axion mass.
More importantly, the axion number density gets fixed when the time
scale of the oscillation, $1/m_a(T)$, is comparable to that of the
change of the axion mass, $(\dot m_a(T) / m_a(T))^{-1}$, so that the
change of the mass is adiabatic, i.e.,
\begin{align}
 m_a(T)& \gtrsim \left| {1 \over m_a(T)} {d m_a(T) \over dt} \right|
= H(T) \left| {d \ln m_a(T) \over d \ln T} \right|.
\label{eq:variation}
\end{align}
The condition is automatically satisfied when $m_a(T) \sim H(T)$ and
$m_a(T) \propto T^n$ with $n \sim O(1)$, but not necessarily true for
exponential functions.

In the model discussed in the previous section, $m_a(T)$ is given by
\begin{align}
 m_a(T) = m_a(T_c) e^{-c (T^2 - T_c^2)/ T_c^2},
\label{eq:assumption}
\end{align}
where the $c$ parameter may be much larger than $O(1)$.
The oscillation starts when both sides of Eq.~\eqref{eq:variation} become
comparable, and thus
\begin{align}
 m_a(T_*) \sim 2 c H (T_*) 
\left( {T_* \over T_c} \right)^2.
\end{align}
Since $m_a(T)$ is a steeply varying function, we expect $T_* \sim T_1$,
where $T_1$ is defined by
\begin{align}
 m_a(T_1) = 3 H (T_1).
\end{align}
This causes an enhancement of the axion density by a factor of about $c
(T_1/T_c)^2$ compared to the conventional estimate in addition to
the enhancement due to the shift of the oscillation temperature.

A numerical solution of the equation of motion in
Eq.~\eqref{eq:axionEOM} shows that the axion number density is
approximately given by
\begin{align}
 {n_a(T) \over T^3} \simeq 
{0.5 c \times 3 H(T_1) f_a^2 \theta^2 \over T_1^3}
\cdot
{T_1^2 \over T_c^2}
=
{0.5 c \times 3 H(T_c) f_a^2 \theta^2 \over T_c^3}
\cdot
{T_1 \over T_c}.
\end{align}
Here
\begin{align}
 T_1 \sim T_c \left[
1
+ {1 \over c} \left(12 + \log
\left(
m_a (T_c) \over 10^{-5}~{\rm eV}
\right)
\right)
\right]^{1/2}.
\label{eq:temp1}
\end{align}
Here we used $3 H(T_c) \simeq 5\times 10^{-20}$~GeV, and assumed
$(T_1/T_c)^2 \ll m_a(T_c) / 3 H(T_c)$.
Putting it all together, the axion energy density today is given by
\begin{align}
\Omega_a &\sim 
0.2 \cdot \theta^2 \cdot
\left(
{m_a \over 10^{-5}~{\rm eV}}
\right)^{-1}
\times
2.5 c
\left[
1
+ {1 \over c} \left(12 + \log
\left(
m_a (T_c) \over 10^{-5}~{\rm eV}
\right)
\right)
\right]^{1/2}.
\label{eq:Omega}
\end{align}
Even for $c = O(1)$ and $m_a = 10^{-5}$~eV, we find that the axion
energy density is enhanced by an order of magnitude compared to the
conventional estimate. 
The enhancement is larger for larger $c$. The axion window is closed
when $c \sim 400$.
For an extremely large value of $c$, i.e., $c \gtrsim 4 \times 10^5
\cdot (m_a / 10^{-5}~{\rm eV} )$, the axion starts to oscillate at $T =
T_c$. In that case, $\Omega_a \sim 2 \times 10^5 \cdot \theta^2$
independent of $c$.
 
We have also examined the case of $\chi_t \propto \exp (-2c\,T/T_c)$
rather than $\chi_t \propto \exp (-2c\,T^2/T_c^2)$ in
Eq.~(\ref{eq:naivemodel2}).
We find a similar enhancement factor $1.0\,c$ instead of 
$2.5\,c$ in Eq.~(\ref{eq:Omega}).

In the scenario where the Peccei-Quinn symmetry breaking takes place
after the inflation, one should take an average value of $\theta$,
$\langle \theta^2 \rangle \simeq \pi^2$.
Also, in addition to the misalignment mechanism considered above, there are
contributions to the axion density from the axionic strings and domain
walls~\cite{Davis:1986xc, Davis:1989nj, Sikivie:1982qv,
Lyth:1991bb}. Both contributions receive an enhancement due to the delay
of time scale of the axion production which leads a milder dilution due
to the cosmic expansion. The enhancement factor of $\Omega_a$ compared
to the estimates based on $m_a(T) \propto T^{-4}$, $\Omega_a^{\rm
conv.}$, is the ratio of the temperature in Eq.~\eqref{eq:tempS} and
the temperature $T_1$ at which the adiabatic condition is satisfied.
The enhancement factor is
\begin{align}
 \Omega_a / \Omega_a^{\rm conv.} \Bigg|_{\rm string+wall} 
\simeq 6.6 \times 
\left(
{m_a \over 10^{-5}~{\rm eV}}
\right)^{1/6}
\left[
1
+ {1 \over c} \left(12 + \log
\left(
m_a (T_c) \over 10^{-5}~{\rm eV}
\right)
\right)
\right]^{-1/2},
\end{align}
For $m_a = 10^{-5}$~eV, for example, the enhancement is a factor of
a few for $c = O(1)$. The value of $\Omega_a^{\rm conv.}|_{\rm string+wall}$ is
found in Ref.~\cite{Kawasaki:2014sqa}:
\begin{align}
 \Omega_a^{\rm conv.} \Big|_{\rm string+wall} 
\sim 4 \times \left(
m_a \over 10^{-5}~{\rm eV}
\right)^{-7/6}.
\end{align}

\section{Summary}

The temperature dependence of the topological susceptibility $\chi_t$
essentially determines the abundance of the QCD axion in the Universe,
and can be calculated using lattice QCD in model independent way.
However, at high temperatures, the lattice calculation suffers from
various difficulties such as the numerically demanding definition,
possible misidentifications and the long autocorrelation of the
topological charge.

As the first step towards the precise calculation, we have performed the
lattice calculation of $\chi_t$ in the $SU(3)$ Yang-Milles theory,
focusing on the autocorrelation time in HMC, and found that the problem
starts to show up around $T \sim 2.0\,T_c$,
This immediately indicates the difficulty in dynamical simulations at
the similar temperature while estimating the axion abundance in the
conventional scenario requires the $T$ dependence up to $O(10\,T_c)$.
To fill this gap, we need exploratory studies of HMC parameters or even
algorithm itself.

Meanwhile, in order to see the importance of understanding the
temperature dependence of $\chi_t$ for the axion abundance in the
Universe, we have calculated the axion energy density, $\Omega_a$, in an
extreme case where $\chi_t$ drops exponentially above the critical
temperature.
It is found that the non-adiabatic growth of the axion potential makes
the axion abundance significantly larger, possibly closing the axion
window.
In this particular case, the behavior of $\chi_t$ around $T_c$ becomes
important.

Our findings provides strong motivation to further serious studies of
topology on the lattice and instanton at medium temperatures.

\section*{Acknowledgments}

We would like to thank Michael Dine for useful conversations and Hideo
Matsufuru for providing us with his lattice codes.
We would also like to thank Evan Berkowitz, Michael Buchoff and Enrico
Rinaldi for useful comments.
This work was supported by JSPS KAKENHI Grant-in-Aid for Scientific Research (B)
(No.~15H03669 [RK, NY]) and MEXT KAKENHI Grant-in-Aid for Scientific
Research on Innovative Areas (No.~25105011 [RK]).

\section*{Note added}

While this paper was being completed, Ref.~\cite{Berkowitz:2015aua}
appeared which estimates $\chi_t$ at high temperatures in the quenched
approximation with the heatbath algorithm on large volumes and uses it
to discuss the axion abundance.
It turns out that our results at $N_t=4$ are consistent with theirs to
$T \sim 2\,T_c$, which give us a more confidence that the
autocorrelation time of topological charge increases with volume in HMC
and hence increasing volume does not resolve the problem.


\end{document}